\documentclass[aps,prb,superscriptaddress,showpacs,onecolumn,12pt]{revtex4}
\usepackage{amsfonts}
\usepackage{amsmath}
\usepackage{amssymb}
\usepackage{graphicx}

\begin{document}

\title{Atomistic simulations of self-trapped exciton formation in silicon
  nanostructures: The transition from quantum dots to nanowires}
\author{Yong Wang}
\affiliation{Bremen Center for Computational Material Science,
University of
  Bremen, 28359 Bremen, Germany}
\affiliation{Institute of Physics, Chinese Academy of Sciences,
P.O.\ Box 603, Beijing, 100190, China}
\author{Ruiqin Zhang}
\affiliation{Centre of Super-Diamond and Advanced Films (COSDAF) and
Department of Physics and Materials Science, City University of Hong
Kong, Hong Kong SAR, China}
\author{Thomas Frauenheim}
\affiliation{Bremen Center for Computational Material Science,
University of
  Bremen, 28359 Bremen, Germany}
\author{Thomas A.\ Niehaus}
\email{thomas.niehaus@bccms.uni-bremen.de}
\affiliation {Bremen
Center for Computational Material Science, University of
  Bremen, 28359 Bremen, Germany}
\begin{abstract}
Using an approximate  time-dependent density functional theory
method, we calculate the absorption and luminescence spectra for
hydrogen passivated silicon nanoscale structures with large aspect
ratio. The effect of electron confinement in axial and radial
directions is systematically investigated. Excited state relaxation
leads to significant Stokes shifts for short nanorods with lengths
less than 2 nm, but has little effect on the luminescence intensity.
The formation of self-trapped excitons is likewise observed for
short nanostructures only; longer wires exhibit fully delocalized
excitons with neglible geometrical distortion at the excited state
minimum.
\end{abstract}
 \maketitle

\section{Introduction}
  Research on silicon nanowires (SiNW) has intensified  in the past years, owing to their
potential applications in future nano-technologies, such as
nanosensors \cite{Cui2001}, nanoscale electronics and photonics
devices.\cite{Cui2001a, Gudiksen2002, Duan2003} This interest
results from the electronic structure of SiNW beeing critically
dependent on  the size, orientation, passivation and doping level of
the nanostructure. For example, it has been shown experimentally
that the band gap can be tuned by choosing different growth
directions and diameters for the wire.\cite{Ma2003}  On the
theoretical side, the ground state electronic structure and
transport properties of different kinds of SiNW have been
investigated by several
authors.\cite{Zhao2005,Zhang2005,Ponomareva2005,Yan2006,Bruno2007,Aradi2007,Rurali2007}

   Light absorption and emission of these systems  have also attracted great attention, because
the band gap in bulk crystalline Si is small and indirect, while
that in SiNW can become large and direct due to the quantum
confinement
effect.\cite{Wolkin1999,Yu1999,Li2002,Lyons2002,Guichard2006} This
experimental progress paves the way to obtain visible light from
silicon materials, and provides for the possibility of future
optoelectronics applications. In the past, theoretical descriptions
of the optical properties of SiNW were mainly restricted to
determining their absorption
spectrum.\cite{Zhao2005,Bruno2007,Aradi2007} To quantify
luminescence, it is often tacitly assumed that  de-excitation occurs
resonantly. However, absorption and emission spectra can differ
considerably due to excited state relaxation, and simulations that
take this Stokes shift into account are currently missing for SiNW.
\section{Method and Simulation Approach}
   In this letter, we use the time-dependent
density-functional based tight-binding (TD-DFTB) method
\cite{Niehaus2001a} (for a recent review see \cite{Niehaus2009}), to
study the excited states and optical properties of SiNW. The same
method has been applied to silicon quantum dots,
\cite{Wang2007,WangZLNF2007,LiZNFL2007,WangZLFN2008,LiZLNF2008,LiZLNF2008_2,LiZNFL2007_2}
where it was found to provide a high degree of reliability and
computational efficiency compared with the parental time-dependent
density functional theory (TD-DFT) from which it is derived. For
example, the optical gap of Si$_{5}$H$_{12}$ is predicted to be 6.40
eV, in very good agreement with the experimental value of 6.50
eV.\cite{Feher1977} Also for larger clusters like Si$_{35}$H$_{36}$,
the TD-DFTB estimate for the lowest allowed singlet transition (4.37
eV) coincides with high level {\em ab-initio} results from
multi-reference second order perturbation theory (4.33
eV).\cite{Zdetsis2006}

In principle, SiNW should be treated as quasi one-dimensional
periodic systems, since their lengths are generally in the
micrometer range while their diameters are several nanometers
only.\cite{Ma2003} However, the TD-DFTB method we use here is
currently restricted to deal with finite systems, thus, we construct
silicon nanostructures along the $\langle 110\rangle$   direction
with increasing length to approach the experimentally realized SiNW.
This allows us to demonstrate how the localized excitons in silicon
quantum dots evolve into delocalized ones
 as the length of the Si nanostructures continously
 increases. Focussing on finite structures also guarantees that we do
 not leave the trust region of TD-DFT. For periodic systems, local or
 semi-local approximations to the exchange-correlation functional lead
 to a collapse of the many-body excited state energies on the ground
 state Kohn-Sham gap.\cite{izmaylov2008tdd,Botti2007} A better description of the electron-hole
 interaction is provided by quasi-particle calculations at the GW
 level combined with the solution of the Bethe-Salpeter equation.\cite{Onida2002,Ramos2008} For
 the confined systems discussed here, however, TD-DFT provides a reliable and
 affordable way to investigate optical properties.

\begin{figure}
\includegraphics[scale=0.35]{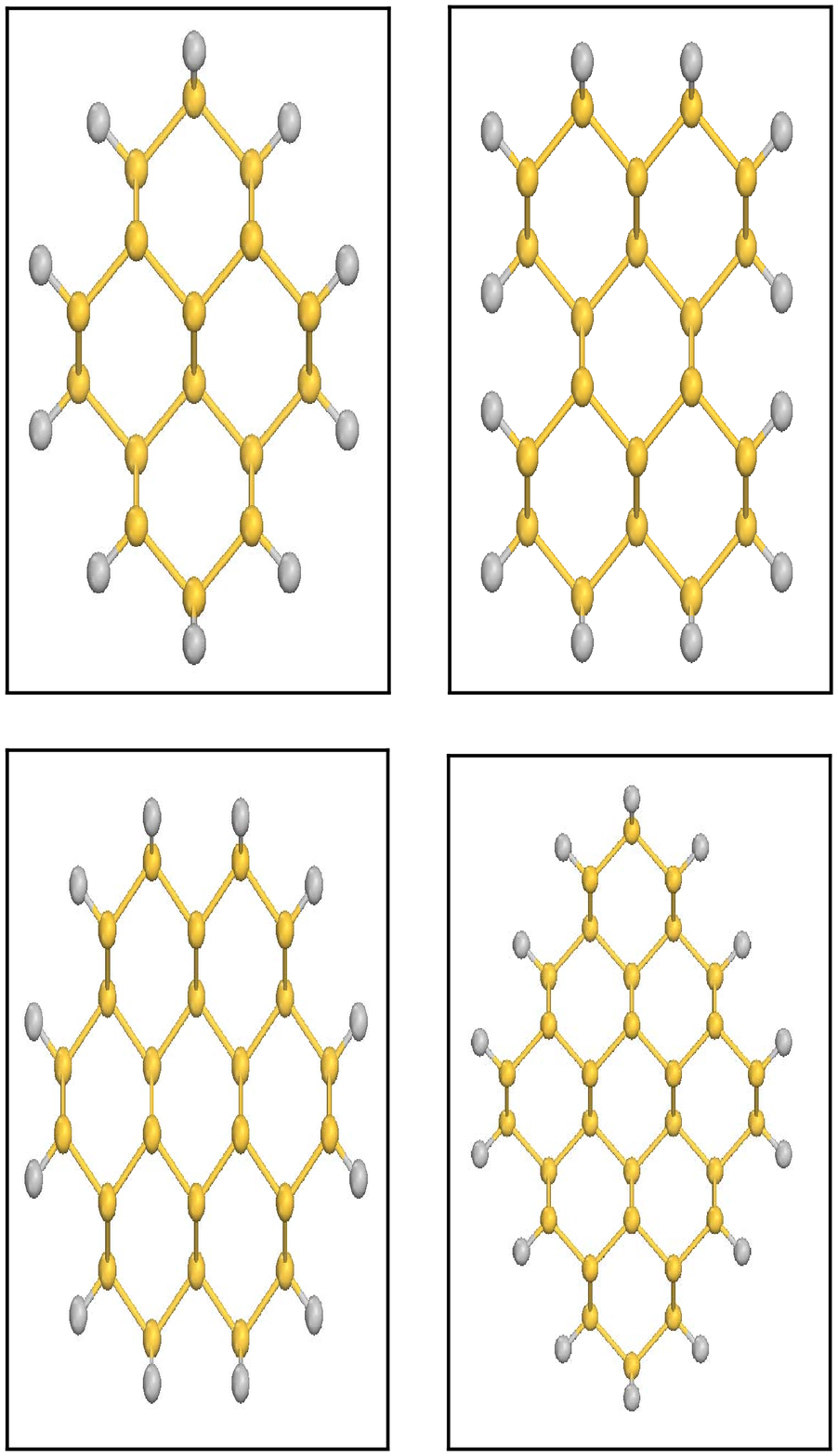}
\caption{(color online) Ball and stick models of four $\langle
110\rangle$  silicon nanowires (SiNW) with different diameters
viewed from the top. Here Si atoms are yellow-colored and H atoms
are grey-colored.}\label{structure}
\end{figure}

  We studied $\langle 110\rangle$  SiNW with four different diameters, as shown in
\ref{structure}. The diameters $d$ are estimated to be 0.84 nm, 0.98
nm, 1.08 nm, 1.24 nm, respectively. Due to the limitation of
computational resources, SiNW with other growth directions or larger
diameters were not considered.
The simulation protocol for each structure can be summarized as
follows: First, to build the structural model, we fully saturated
the dangling bonds of surface Si atoms with hydrogen atoms. Second,
the constructed models were relaxed in the ground state by conjugate
gradient optimization using the DFTB method.\cite{elstner1998scc}
Third, we obtained the low energy part of the absorption spectrum at
the ground state optimum by solving for the energies and oscillator
strengths for the lowest five singlet excitations.  And fourth,
assuming rapid internal conversion, (i.e.~the validity of Kasha's
rule), the emission energies are evaluated by geometry optimization
in the first excited singlet state (S$_1$). For most systems studied
here, this assumption does not represent an approximation, since the
strongest absorbing state is actually the S$_1$. In the optimization
process, we took advantage of a recent implementation of analytical
excited state forces for the TD-DFTB scheme.\cite{heringer2007aes}
In all calculations the basis set consists of $s$ and $p$ orbitals
for Si atoms and an $s$ orbital for H atoms, and the
gradient-corrected PBE exchange-correlation functional is
employed\cite{Note-1}.

  For a given structure, excited state energies ($\omega_I$) are obtained in two steps.\cite{Niehaus2001a,Niehaus2009}
First, ground state spin-restricted DFTB calculations are performed
to obtain the Kohn-Sham (KS) orbitals $\psi_{i}$ and the KS energies
$\epsilon_{i}$. These single-particle values are corrected following
the TD-DFT linear response treatment of Casida\cite{Casida1995}
  \begin{equation}
  \label{excite}
  \sum_{kl\tau}[\omega_{ij}^{2}\delta_{ik}\delta_{jl}\delta_{\sigma\tau}
  +2\sqrt{\omega_{ij}}K_{ij\sigma,kl\tau}\sqrt{\omega_{kl}}]F_{kl\tau}^{I}=\omega_{I}^{2}F_{ij\sigma}^{l},
  \end{equation}
where $\omega_{ij}=\epsilon_{j}-\epsilon_{i}$ and $\sigma$ and
$\tau$ are spin indices. The coupling matrix $K_{ij\sigma,kl\tau}$
defined as\cite{Casida1995}
  \begin{eqnarray*}
  K_{ij\sigma,kl\tau}&=&\int\int\psi_{i}(\textbf{r})\psi_{j}(\textbf{r})
  (\frac{1}{|\textbf{r}-\textbf{r}'|}+\frac{\delta^{2}E_{xc}}
  {\delta\rho_{\sigma}(\textbf{r})\delta\rho_{\tau}(\textbf{r}')})\\
  &&\times\psi_{k}(\textbf{r}')\psi_{l}(\textbf{r}')
  d\textbf{r}d\textbf{r}',
  \end{eqnarray*}
  is further simplified in the TD-DFTB approach using the Mulliken approximation.\cite{Niehaus2001a,Niehaus2009}

  With the results of equation (\ref{excite}), oscillator strengths are calculated as \cite{Casida1995}
  \begin{equation}
  \label{osci}
  f_{I}=\frac{2}{3}\omega_{I}\sum_{k=x,y,z}|\sum_{ij}<\psi_{i}|\textbf{r}_{k}|\psi_{j}>
  \sqrt{\frac{\omega_{ij}}{\omega_{I}}}(F_{ij\uparrow}^{I}+F_{ij\downarrow}^{I})|^{2}
  \end{equation}

\section{Results and Discussion}
\begin{figure}
\centering
\includegraphics[scale=0.4, angle=270]{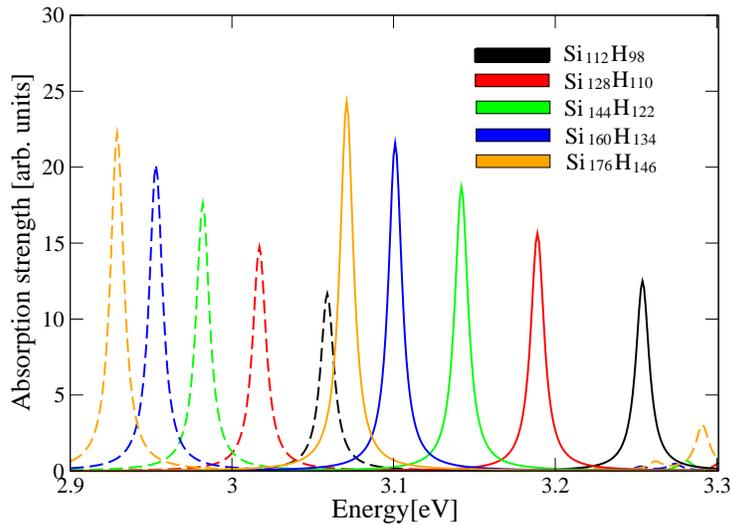}
\caption{(color online) Absorption (solid lines) and emission
spectra (dashed lines) for $\langle 110\rangle$   nanowires with
diameter 0.84 nm and different lengths (from 2.7 nm to 4.3 nm). The
spectra have been broadened by 0.01 eV to simulate finite
temperature. }\label{spectra}
\end{figure}

  In \ref{spectra}, we show the absorption and luminescence spectra
for the thinnest $\langle 110\rangle$  nanowire
($d=0.84\text{\,nm}$) with different lengths, from $l=2.7nm$
($Si_{112}H_{98}$) to $l=4.2nm$ ($Si_{176}H_{146}$). Similar to what
we have previously observed in silicon quantum dots with increasing
diameter,\cite{Wang2007,WangZLNF2007}
 we find here that both absorption and emission
energies slightly red-shift with increasing wire length. Moreover, a
sizable Stokes shift of around $0.1\sim 0.2$ eV is discernible.
Since excited state relaxation does not lead to a significant change
in oscillator strengths, all considered nanostructures should
exhibit significant luminescence intensity.

\begin{figure}[tbp]
\includegraphics[scale=0.6, angle=270]{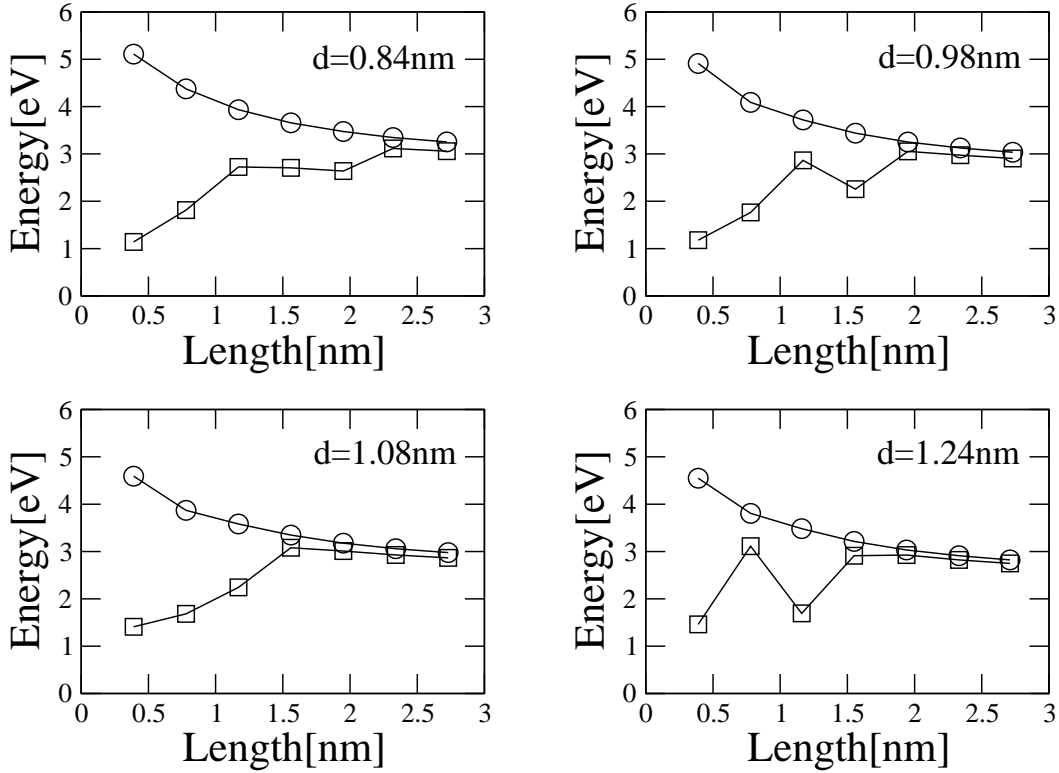}
\caption{TD-DFTB absorption energies (circles) and emission energies
(squares) for $\langle 110\rangle$   SiNW with different diameter
and lengths. }\label{gapvslen}
\end{figure}

  Next, we extracted the energies of the first allowed transition for SiNW with different diameters to investigate the size dependence of the optical gap (see \ref{gapvslen}).
When the length of the Si nanostructures increases from 0.39 nm to
2.73 nm, the absorption energies $E_\mathrm{abs}$ will decrease
monotonously from  around 5 eV to  around  3 eV. This is due to the
decrease of quantum confinement effects. Concurrently, the emission
energies $E_\mathrm{emi}$ show an overall increase (although with
some fluctuations for most cases) from 1 eV to around 3 eV, and then
follow the trend of the absorption energies. This general behaviour
has also been found in similar studies on the size dependence of the
optical gap in spherical Si quantum
dots.\cite{Wang2007,WangZLNF2007} Quantum confinement effects are
also observed in the radial dimension. Considering a fixed length of
$l=1.56\text{\,nm}$, the optical gap decreases from 3.66 eV for the
thinnest diameter ($d=0.84\text{\,nm}$) to 3.21 eV for
$d=1.24\text{\,nm}$.

  In order to understand the transition from quantum dots to wires more deeply, we analyze the geometrical distortions arising from the excited state relaxation.  In small quantum dots, one Si-Si bond
is extremely stretched  up to  2.70 \AA.\cite{WangZLNF2007}
Depending on the actual structure, the location of this bond can be
either in the center of the cluster or at its surface. When the rods
are long enough, the geometry distortions are small and more
homogenously distributed over all bonds. For the thinnest
nanostructures ($d=0.84\text{\,nm}$), the distortions locate at one
Si-Si bond in the center of the structures for lengths less than
1.95 nm and delocalize beyond this value, owing to the confinement
provided by the rigid surrounding layer.\cite{WangZLNF2007}  For a
diameter of 1.08 nm, geometry changes are observed for one Si-Si
bond at the surface of the structures up to a length of 1.17 nm
($Si_{72}H_{68}$). The situation for the other two cases
($d=0.98\text{\,nm}$ and $d=1.24\text{\,nm}$) is more involved. For
the wire with $d=0.98\text{\,nm}$, distortions are found primarily
for a Si-Si bond at the center of the quantum dots below 0.78 nm
($Si_{40}H_{48}$). The distortion is then delocalized for the
cluster $Si_{60}H_{64}$ at 1.17 nm, which corresponds to the peak in
the emission energy curve in \ref{gapvslen}. It should be mentioned
here that our simulation protocol locates the local minimum closest
to the the Frank-Condon point, i.e.\ the initial structure after
light absorption. It cannot be ruled out that the global minimum of
the $S_1$ potential energy surface is again of localized nature.
Whether luminescence occurs also from this state will crucially
depend on the excited state lifetime and the kinetic energy of the
ions after absorption. Excited state molecular dynamics simulations
could shed light on this interesting question, which is however
outside the scope of this study.

Turning back to the results, a surface Si-Si bond is found to be
elongated for the cluster $Si_{80}H_{80}$ at 1.56 nm and thereafter
all distortions have a delocalized character for longer structures.
A similar case are the wires with $d=1.24\text{\,nm}$, where a
delocalized distortion appears for $Si_{60}H_{60}$
($l=0.78\text{\,nm}$), and a localized distortion for
$Si_{90}H_{76}$ ($l=1.16\text{\,nm}$).

\begin{figure}
\includegraphics[scale=0.7]{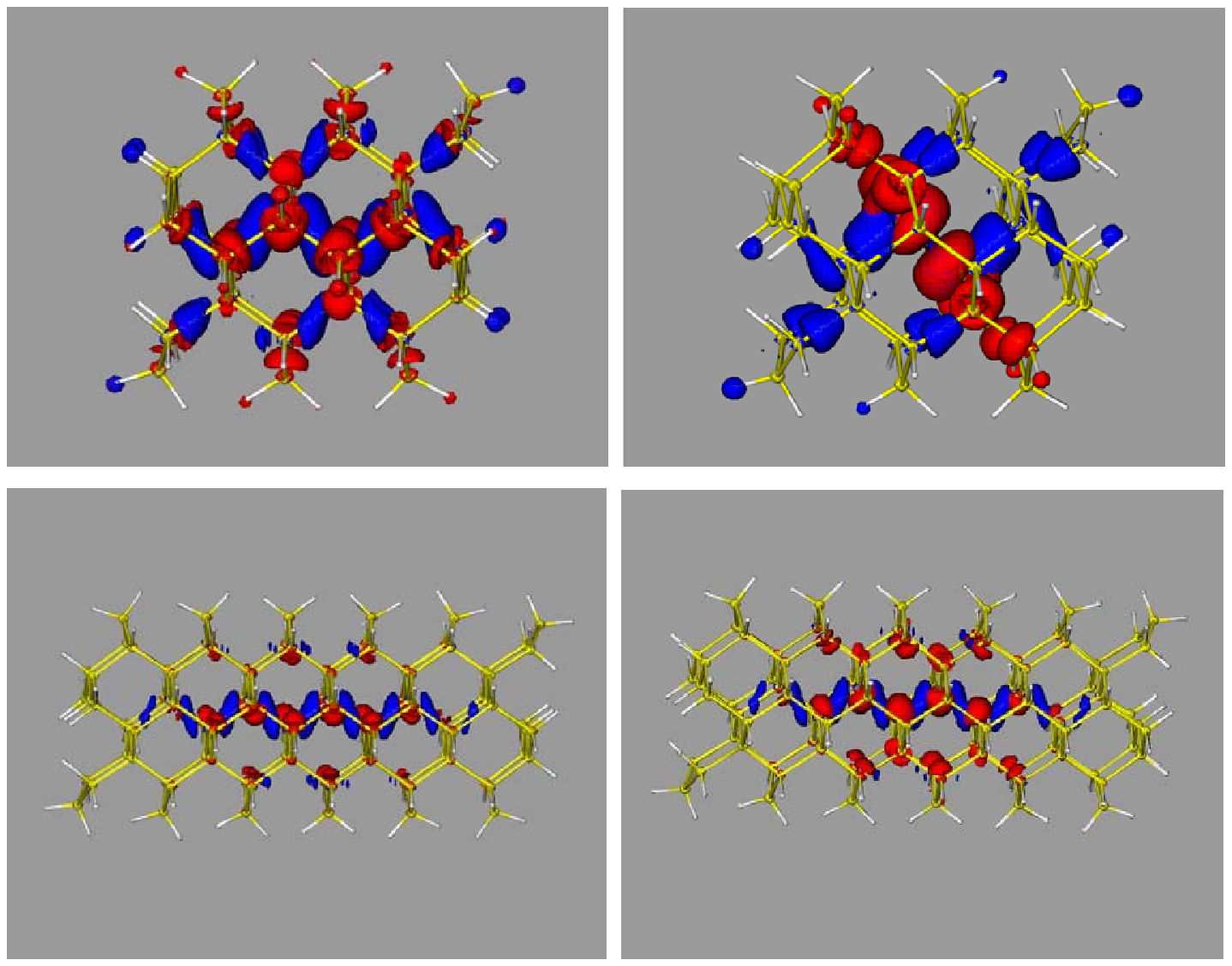} \caption{(color online)
Absolute values of the molecular orbitals HOMO (blue) and LUMO (red)
for ground state optimized (left) and excited state optimized
(right) structures of $Si_{48}H_{50}$ (up) and $Si_{96}H_{86}$
(down), which are models of $\langle 110\rangle$   SiNW with
diameter $d=0.84\text{\,nm}$. The plot corresponds to an isovalue of
0.001.}\label{HomoLumo}
\end{figure}

  Additional information on the electronic structure can be extracted
  from the KS molecular orbitals that are involved in the light
  absorption and emission. \cite{Note-2} We take the rods
$Si_{48}H_{50}$ ($l=1.17\text{\,nm}$) and $Si_{96}H_{86}$
($l=2.32\text{\,nm}$) with diameter $d=0.84$ as examples. In both
systems, the $S_1$ excited state wavefunction is dominated by a
single-particle transitions from the highest occupied molecular
orbital (HOMO) to the lowest unoccupied molecular orbital (LUMO). In
\ref{HomoLumo}, these orbitals are depicted for the optimized
geometries in $S_0$ and $S_1$ . We find for the ground state minimum
of both structures that the  HOMO  distributes mainly along the
central zone of the wire, while the LUMO  is located on the surface.
The molecular orbitals at the excited state minimum conformation are
quite different for the two structures under investigation. The
shorter structure, $Si_{48}H_{50}$, exhibits a LUMO concentrated at
the center of the cluster that gives rise to a repulsive force. The
significant Si-Si bond stretch in combination with the resulting
sizable Stokes shift is completely in line with the self-trapped
exciton model of Allan, Delerue, and
Lannoo.\cite{allan1996nls,Note-3} In contrast, the LUMO of the
longer rod, $Si_{96}H_{86}$,  distributes more or less homogenously
along the structure, indicating a delocalized exciton and geometry
distortion.

\section{Concluding Remarks}
In conclusion, we simulated the absorption and emission spectra of
SiNW with different diameters. The evolution of the optical
properties from small quantum dots to nanowires with large aspect
ratio has been investigated. While short nanorods with lengths below
2 nm show localized excitations and the formation of self-trapped
excitons, excited state relaxation has little effect on longer
structures which exhibit delocalized excited states. Notwithstanding
these general trends, silicon nanostructures with similar extensions
can exhibit quite different localization characteristics conjoined
with  largely differing emission profiles. This fact may partially
explain the reported spread of photoluminescence energies
\cite{wilcoxon1999oae} and calls for atomistic simulations of these
systems which take this strong conformation dependence into account.

\section{acknowledgement} Yong Wang thanks Dr.\ Xian Wang and Dr.\
Binghai Yan
 for useful discussions and MOST (No. 2006CB933000) of
China for financial support.

\end{document}